\documentclass[aps,prc,twocolumn,superscriptaddress,10pt,english,preprintnumbers, nofootinbib]{revtex4-1}
\pdfoutput=1
\usepackage[T1]{fontenc}
\usepackage[utf8]{inputenc}
\usepackage{epsfig}
\usepackage{graphicx}
\usepackage{xcolor}
\usepackage{latexsym}
\usepackage{textcomp}
\usepackage{amssymb}
\usepackage[colorlinks=true,linkcolor=blue,citecolor=blue, urlcolor=blue]{hyperref} 
\usepackage{amsfonts,amsthm,amstext,amscd}
\usepackage{amsmath}
\usepackage{mathtools}

\def\npart{$N_{\rm part}$}
\def\ncoll{$N_{\rm coll}$}
\def\zr{$^{96}_{40}$Zr}
\def\ru{$^{96}_{44}$Ru}

\begin{document}
\title{Hard probes in isobar collisions as a probe of the neutron skin}
\author{Wilke van der Schee}
\affiliation{Theoretical Physics Department, CERN, CH-1211 Gen\`eve 23, Switzerland}
\affiliation{Institute for Theoretical Physics, Utrecht University, 3584 CC Utrecht, The Netherlands}
\author{Yen-Jie Lee}
\affiliation{Massachusetts Institute of Technology, Cambridge, MA 02139, USA}
\author{Govert Nijs}
\affiliation{Center for Theoretical Physics, Massachusetts Institute of Technology, Cambridge, MA 02139, USA}
\author{Yi Chen}
\affiliation{Massachusetts Institute of Technology, Cambridge, MA 02139, USA}
\begin{abstract}
We present an estimate of the yield of hard probes expected for collisions of the isobars  $^{96}_{44}$Ru and $^{96}_{40}$Zr at collision energies reachable at RHIC and the LHC\@. These yields are proportional to the number of binary nucleon-nucleon interactions, which is characteristically different due to  the presence of the large neutron skin  in $^{96}_{40}$Zr. This provides an independent opportunity to measure the difference between the neutron skin of $^{96}_{44}$Ru and $^{96}_{40}$Zr, which can provide an important constraint on the Equation of State of cold neutron-rich matter.
\end{abstract}

\preprint{CERN-TH-2023-140//MIT-CTP/5588}

\maketitle

\section{Introduction}

The relativistic heavy ion programs at RHIC and the LHC~\cite{Citron:2018lsq,Arslandok:2023utm} aim to extract properties of the quark-gluon plasma (QGP), a state of matter believed to have existed in the early universe~\cite{Busza:2018rrf}\@. However, current knowledge of the initial condition of the QGP, especially how it is formed and shaped from the colliding nuclei, remains limited. Isobar collisions \cite{STAR:2021mii}, involving nuclei with significant differences in structural properties but similar size, offer a new way to study the QGP\@. In particular, it is possible to study ratios of observables obtained from collisions of different isobars, such as the momentum anisotropies of detected particles or the total multiplicity as a function of impact parameter (centrality) and the transverse momenta of the particles. These ratios will have significantly reduced theoretical and experimental systematic uncertainties.

Given the similarity in size between both isobars, the ratio of observables becomes relatively insensitive to shared properties, such as the speed of sound or shear viscosity. However, it becomes particularly sensitive to differences in the shape of the isobars \cite{Zhang:2021kxj, Xu:2021uar, Nijs:2021kvn, Luzum:2023gwy}\@. Subsequently, a precise understanding of the shape contributes to minimising uncertainties in determining QGP properties through data analysis.

The nuclear structure program aims to explain the emergence of nuclei from fundamental theory \cite{Hergert:2020bxy,Ekstrom:2022yea, Hu:2021trw}\@. Synergising with the hot QCD program based on high-energy heavy-ion collisions, this field can benefit from event-by-event measures of particle angular correlations in the final stages of such collisions \cite{Bally:2022vgo, Giacalone:2023cet}. These angular correlations are sensitive to the many-body distribution and correlations of nucleons, including deformations, in the colliding nuclei. High-energy colliders are thus a novel tool for gaining insight into strongly correlated atomic nuclear systems and testing ab initio nuclear structure calculations.

In this paper, we present a study of the yield of hard probes expected for the collision of isobars $^{96}_{44}$Ru and $^{96}_{40}$Zr at energies for RHIC and the LHC\@. Hard probes generically can be separated into colour-neutral probes (such as photons and W and Z bosons) that do not interact with QGP and coloured probes such as quarks and gluons (partons). Due to the transparency in the QGP the former can be used to directly access the production of hard probes, while quarks and gluons can be used to (also) study the interaction with the QGP\@. We will show that the production is significantly larger for $^{96}_{44}$Ru due to its smaller and denser nucleus. Furthermore, while $^{96}_{40}$Zr is larger, the lower temperature has a compensating effect, and only subleading effects are expected for the difference in the interactions of quarks and gluons with the QGP\@.
This approach provides an independent opportunity to measure the difference between the neutron skins of $^{96}_{44}$Ru and $^{96}_{40}$Zr.%

\begin{figure*}[ht]
\includegraphics[width=0.9\textwidth]{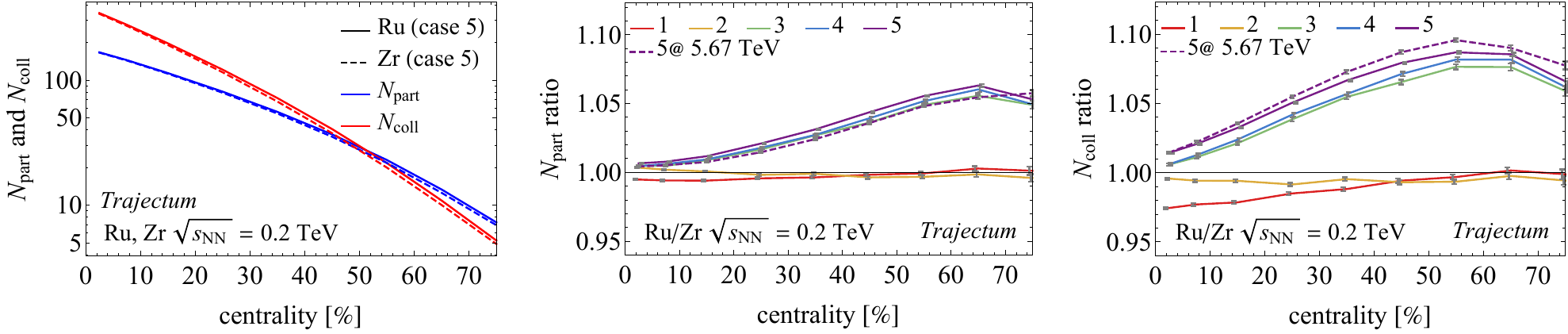}
\caption{\label{fig:ncoll} (Left) The number of participants $N_{\rm part}$ (blue) and number of binary collisions $N_{\rm coll}$ (red) as a function of collision event centrality in RuRu (solid) and ZrZr (dashed) collisions at $\sqrt{s_{NN}}=0.2\,$TeV  for the parameters of Case 5 in Table~\ref{tab:WSparameters}\@. (Middle and Right) The ratios of $N_{\rm part}$ (middle) and $N_{\rm coll}$ (right) in RuRu and ZrZr collisions as a function of centrality for the five cases as presented in Table~\ref{tab:WSparameters}\@. The number of binary collisions is more sensitive to the nuclear structure than the number of participants. The dashed curve presents the same ratio for collisions at a nucleon-nucleon energy of 5.67 TeV, which gives larger effects for $N_{\rm coll}$ and smaller effects for $N_{\rm part}$\@.
}
\end{figure*}

\section{Production of hard probes}

At large transverse momentum the heavy ions are to an excellent approximation transparent, which means that the production rate is proportional to the total number of nucleon-nucleon collisions, i.e.~the number of binary collisions \ncoll{} \footnote{In this work we neglect the modification of the nuclear parton distribution functions (nPDFs)\@. For many of the isospin symmetric probes we expect similar modifications for \ru{} and \zr{}, such that for ratios their effect is reduced.}. The interactions, to be more fully described later, will also depend on the QGP formation and its evolution. Here we describe the model used, with a particular focus on the shapes of the nuclei, since as discussed this has a large effect on the ratios.

We use the \emph{Trajectum} 1.2 framework \cite{Nijs:2020ors, Nijs:2021kvn}\footnote{The \emph{Trajectum} code can be found at \url{https://sites.google.com/view/govertnijs/trajectum}.} using the maximum likelihood settings as in \cite{Nijs:2021clz}\@.
\emph{Trajectum} features an initial state depending on the nucleon positions that generalises the T\raisebox{-0.5ex}{R}ENTo model \cite{1605.03954}, a far-from-equilibrium stage that can interpolate between free streaming and a holographic scenario \cite{Nijs:2023yab}, a hydrodynamic phase with temperature-dependent first and second order transport coefficients and finally a freeze-out prescription \cite{Bernhard:2018hnz} that can transfer the resulting QCD resonance states to a hadronic rescattering code such as UrQMD \cite{nucl-th/9803035} or SMASH \cite{Weil:2016zrk}.

In the T\raisebox{-0.5ex}{R}ENTo model the nucleon positions are located according to a Woods-Saxon (WS) distribution,
\begin{equation}
\label{eq:roftheta}
P_{p/n}(r,\theta) \propto \left(1 + \exp\left(\frac{r - R(\theta)}{\sigma_{p/n}}\right)\right)^{-1},
\end{equation}
where $R(\theta) = R_{p/n} \cdot \left(1 + \beta_2Y_2^0(\theta) + \beta_3Y_3^0(\theta)\right)$, with $Y_n^0$ the spherical harmonics, the radius $R$, skin depth $\sigma$, quadrupole deformation $\beta_2$ and octupole deformation $\beta_3$\@.
Importantly, especially \zr{} has a large neutron excess which leads to significantly more neutrons at the edge of the nucleus, i.e.~a neutron skin. This is reflected in a larger skin depth of \zr{} as compared to \ru{} \cite{Trzcinska:2001sy}\@. In addition, \zr{} is a strongly deformed nucleus that has a large $\beta_3$ deformation \cite{Zhang:2021kxj, Nijs:2021kvn, Rong:2022qez}, though this deformation plays a relatively minor role for hard probe studies as compared to the neutron skin.

In this work we use the five different parameterisations as reproduced in Table~\ref{tab:WSparameters} from \cite{Nijs:2021kvn}\@. The first three are taken from the STAR isobar paper \cite{STAR:2021mii}, whereby only the third includes the neutron skin effect as determined from Density Functional Theory (DFT)\@. Case 4 comes from the same DFT framework as Case 3, albeit with a fixed elliptic deformation of $\beta_2 = 0.16$ for both \ru{} and \zr{}. Case 5 is finally the most realistic case \cite{Zhang:2021kxj}, with the same neutron skin as Case 4 but including non-equal spherical deformations for \ru{} and \zr{}. Here $\beta_2$ was derived from excitation energies measured in \cite{Pritychenko:2013gwa} and $\beta_3$ was fitted to the STAR result \cite{STAR:2021mii} itself.

Having specified how nucleon positions are sampled, every pair of nucleons interacts with a probability based on their overlap, in such a way that the nucleon-nucleon cross-section $\sigma_{NN}$ equals the proton-proton cross-section $\sigma_{pp}$ for that particular collision energy \cite{Moreland:2018gsh}\@. %
The total number of interactions equals \ncoll{}, whereas the number of nucleons that have at least one interaction is called \npart{}.

In Fig.~\ref{fig:ncoll} we show \npart{}, \ncoll{} and their ratios for \ru{} and \zr{} for the five cases. The cases with neutron skin (3 to 5) lead to a smaller size for \ru{}  and hence a smaller cross section (see Table~\ref{tab:WSparameters})\@. Per collision, however, there are then more participating nucleons (and hence also a higher multiplicity, see \cite{STAR:2021mii, Nijs:2021kvn})\@. For \ncoll{} this effect is significantly stronger, which will give rise to the characteristic signal for hard probes.

For case 5 we also include a collision energy appropriate for the LHC ($5.67\,$TeV per nucleon pair, dashed), which for this figure equals to an increase of $\sigma_{\rm NN}$ from 39.7 to $68.8\,$mb. This increases both \npart{} and \ncoll{}, but interestingly the Ru/Zr ratio decreases for \npart{} while it increases for \ncoll{}\@. This indicates that the \ncoll{} effect will be stronger at higher collision energies.

\begin{table}[ht]
\begin{tabular}{cccccccc}
\hline
\hline
nucleus & $R_p\,$[fm] & $\sigma_p\,$[fm] & $R_n\,$[fm] & $\sigma_n\,$[fm] & $\beta_2$ & $\beta_3$ & $\sigma_\text{AA}\,$[b] \\
\hline
$_{44}^{96}$Ru(1) & 5.085 & 0.46 & 5.085 & 0.46 & 0.158 & 0 & 4.628 \\
$_{40}^{96}$Zr(1) & 5.02 & 0.46 & 5.02 & 0.46 & 0.08 & 0 & 4.540 \\
\hline
$_{44}^{96}$Ru(2) & 5.085 & 0.46 & 5.085 & 0.46 & 0.053 & 0 & 4.605 \\
$_{40}^{96}$Zr(2) & 5.02 & 0.46 & 5.02 & 0.46 & 0.217 & 0 & 4.579 \\
\hline
$_{44}^{96}$Ru(3) & 5.06 & 0.493 & 5.075 & 0.505 & 0 & 0 & 4.734 \\
$_{40}^{96}$Zr(3) & 4.915 & 0.521 & 5.015 & 0.574 & 0 & 0 & 4.860 \\
\hline
$_{44}^{96}$Ru(4) & 5.053 & 0.48 & 5.073 & 0.49 & 0.16 & 0 & 4.701 \\
$_{40}^{96}$Zr(4) & 4.912 & 0.508 & 5.007 & 0.564 & 0.16 & 0 & 4.829 \\
\hline
$_{44}^{96}$Ru(5) & 5.053 & 0.48 & 5.073 & 0.49 & 0.154 & 0 & 4.699 \\
$_{40}^{96}$Zr(5) & 4.912 & 0.508 & 5.007 & 0.564 & 0.062 & 0.202 & 4.871 \\
\hline
\hline
\end{tabular}
\caption{\label{tab:WSparameters}Woods-Saxon parameters and inelastic nucleus-nucleus cross sections for the five cases taken from \cite{Nijs:2021kvn}, for both $_{44}^{96}$Ru and $_{40}^{96}$Zr\@. The $p$ and $n$ labels denote the different WS distributions used for protons and neutrons, respectively.}%
\end{table}

\begin{figure}[ht]
\includegraphics[width=0.5\textwidth]{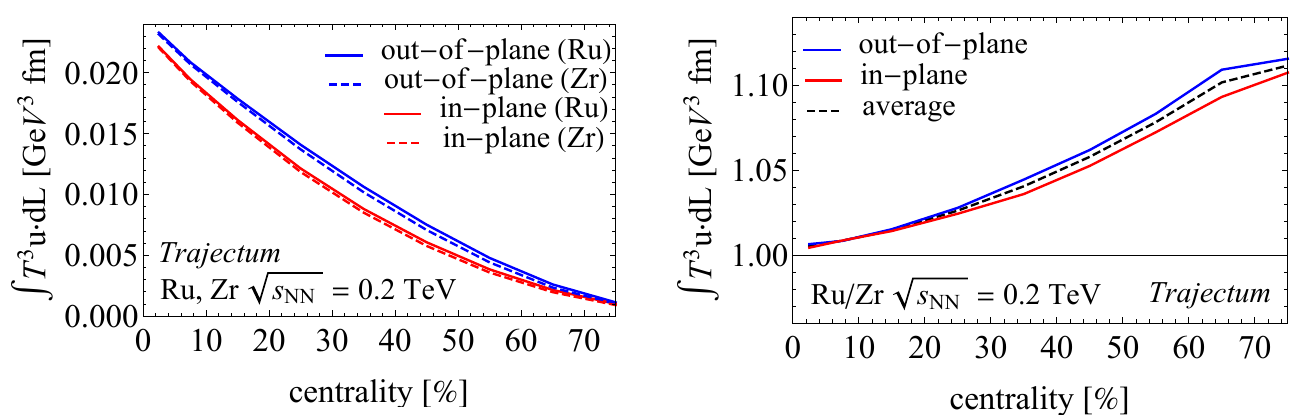}
\includegraphics[width=0.5\textwidth]{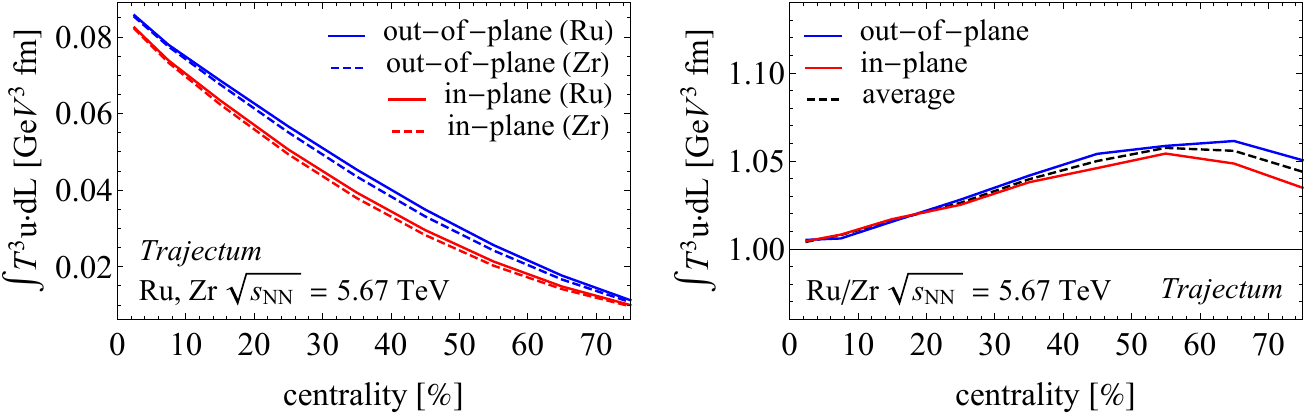}
\caption{\label{fig:pathlengths} (Left) The integrated path lengths calculated in RuRu and ZrZr collisions at $\sqrt{s_{NN}}=0.2\,$TeV (top) and $5.67\,$TeV (bottom)\@. Their ratios are shown in the right panels. The in-plane (out-of-plane) results are shown as red (blue) lines while the average is presented as black lines.
}
\end{figure}

\begin{figure}[h]
\includegraphics[width=0.5\textwidth]{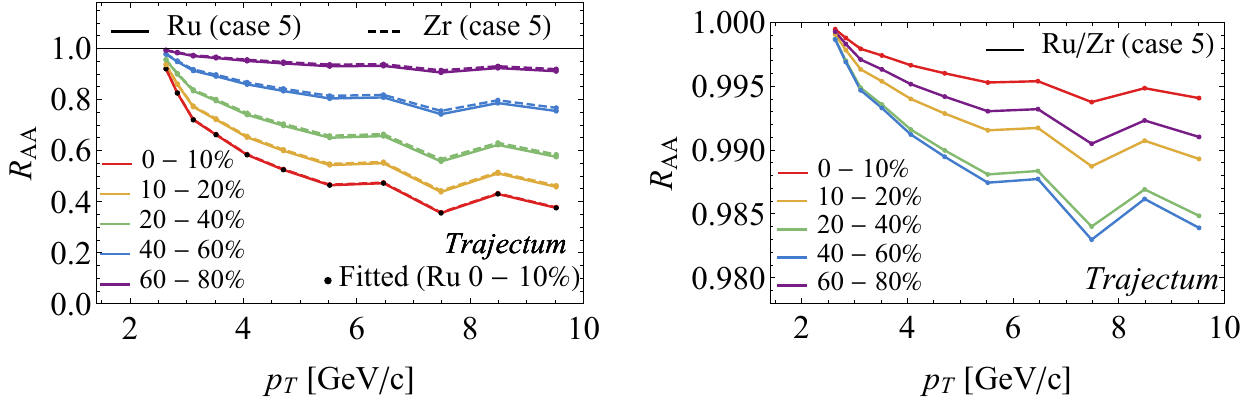}
\caption{\label{fig:RAAratios} We show the charged particle $R_{\rm AA}$ as a function of particle $p_{T}$ in different centrality intervals (left). The Ru results in the 0--10\% class are fitted \cite{STARquarkmatter} and the rest are computed using Eqn.~\ref{eq:raa}\@. The $R_{\rm AA}$ Ru/Zr ratios are shown in the right panel. The ratio is close to unity and is maximal for the 40--60\% centrality class, which optimizes the larger difference in path length (see Fig.~\ref{fig:pathlengths}) while still having an $R_{\rm AA}$ that is significantly different from unity.%
}
\end{figure}

\section{Interactions with the QGP}

After having discussed the production of hard probes we estimate the modification of the hard probes due to the interactions with the QGP\@. 
Due to energy loss of the partons the yield of charged hadrons at high $p_T$ will be reduced. Typically this is quantified using the nuclear modification factor $R_{\rm AA}$, which gives the ratio between the yield in AA collisions versus the expected yield from $pp$ collisions in the absence of a medium such as the QGP\@. For the $R_{\rm AA}$ we estimate that (see also \cite{Beattie:2022ojg})
\begin{eqnarray}
R_{\rm AA}(p_T) &=& \sigma_{pp}(p_T + \delta e (p_T)) / \sigma_{pp}(p_T),    \label{eq:raa}
\\
\delta e (p_T) & = & \kappa(p_T) \int T^3 \bf{u} \cdot \bf{dL},
\end{eqnarray}
where $\delta e (p_T)$ is an estimate for the effective energy loss, $\sigma_{pp}(p_T)$ is the relevant cross section in $pp$ collisions which we estimate as being proportional to $p_T^{-7}$ and $\kappa$ is a $p_T$ dependent constant that we fix from the measurement in the 0--10\% Ruthenium centrality class.
$T$ and $\bf{u}$ are the temperature and fluid velocity a parton encounters\footnote{We average over many events and many lightlike trajectories placed according to the distribution of nucleon-nucleon interactions.} during its path from that start of hydrodynamics at  $\tau = 1.17\,$fm$/c$ till freeze-out at a temperature of $T_\text{switch} = 153.5\,$MeV\@. The motivation for this formula is that energy loss is on average approximately proportional to the entropy density (which goes like $T^3$ for a scale invariant theory) and the $\bf{u} \cdot \bf{dL}$ partly corrects for the fluid flow \cite{Baier:2006pt, Beattie:2022ojg}\@. 

If $\delta e$ is small then $R_{\rm AA}$ is close to unity and can be approximated as\footnote{In the figures we use the exact formula.} 
\begin{equation}
    R_{\rm AA}(p_T) = 1-\frac{\left[1 - R_\text{AA, Ru, 0--10\%}(p_T)\right]\int T^3 u \cdot dL  }{\int T^3 u \cdot dL_{\text{Ru, 0--10\%}}},
\end{equation}
where the subscript Ru 0--10\% indicates that we use the $R_{\rm AA}$ in the 0--10\% class for Ruthenium to fix $\kappa(p_T)$\@. We note that this is a relatively coarse estimate for the charged hadron $R_{\rm AA}$ that in particular ignores fluctuations in energy loss, modification of parton distribution functions or effects coming from fragmentation. Nevertheless, for a comparison between \zr{} and \ru{} we expect that many such effects mostly cancel in the ratio for our observables.

\begin{figure}[hb]
\includegraphics[width=0.5\textwidth]{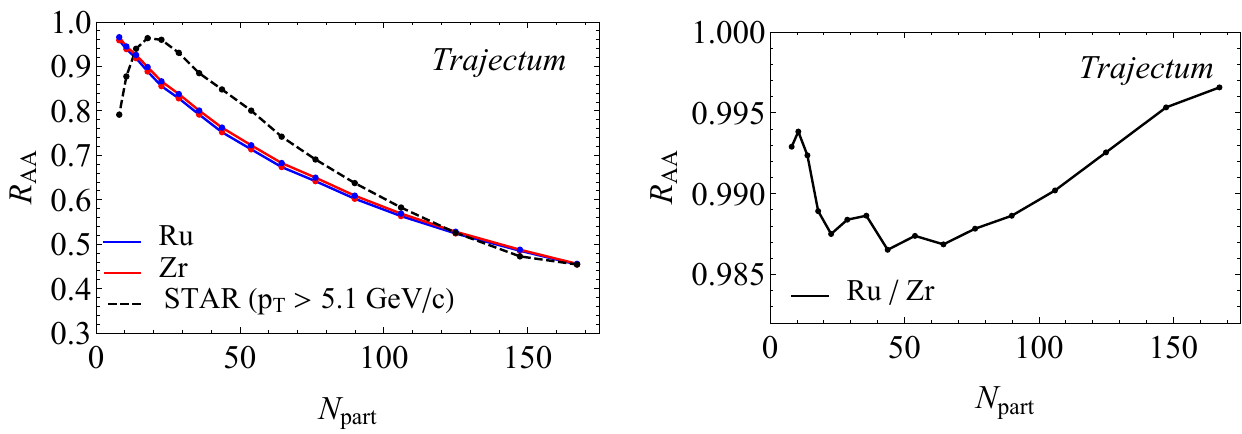}
\caption{\label{fig:RAAvsNpart} (Left) Nuclear modification factor for high $p_{T}$ charged particles ($p_{T}>5.1\,\text{GeV}/c)$) in RuRu (blue line) and ZrZr (red line) compared to STAR RuRu data (black dots, \cite{STARquarkmatter}) as a function of $N_{\rm part}$\@. Again the Ru 0--10\% class is fitted, which corresponds to the rightmost Ru data point. (Right) The ratio of RuRu and ZrZr $R_{\rm AA}$ as a function of $N_{\rm part}$ in \emph{Trajectum}.
}
\end{figure}

\begin{figure*}[ht]
\includegraphics[width=0.95\textwidth]{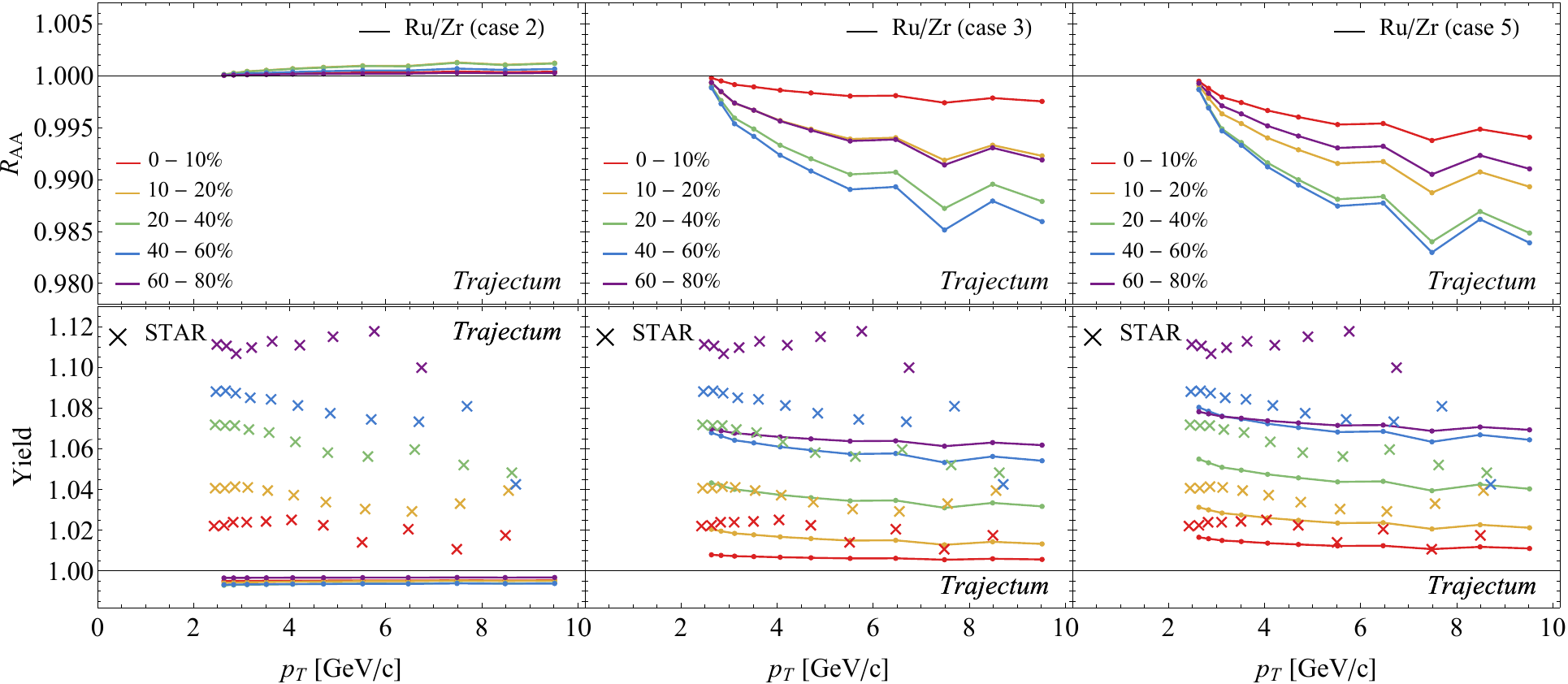}
\caption{\label{fig:yields}For three representative cases (2, 3 and 5 in left, middle and right respectively) we present the $R_{\rm AA}$ ratio between Ru/Zr (top) and the total yields (bottom) as a function of $p_T$ for several centrality classes. As in Fig.~\ref{fig:RAAvsNpart} the centrality dependence in the STAR data \cite{STARquarkmatter} is somewhat stronger in data than in the simple estimate of Eqn.~\ref{eq:raa}\@. The yields are simply a multiplication of the $R_{\rm AA}$ and the number of binary collisions from Fig.~\ref{fig:ncoll}\@. Nevertheless, we argue the yields are a simpler and more sensitive probe of nuclear structure than $R_{\rm AA}$\@.
}
\end{figure*}

\section{Results}

In Fig.~\ref{fig:pathlengths} we present the average temperature integrals %
of in-plane and out-of-plane\footnote{In- and out-of-plane is defined with respect to the $Q_2 = \sum_{i=1}^{M} e^{2 i \varphi_i}$ direction whereby the sum is over all charged particles.} for \ru{} and \zr{} at 0.2 and $5.67\,$TeV\@. It is interesting that even though \ru{} is smaller the temperature is higher and hence the total temperature integral is larger than for \zr{}. This effect is less pronounced at $5.67\,$TeV since the lifetime of the QGP is longer and hence the finite size is increasingly important.

With the temperature integrals, Eqn.~\ref{eq:raa} and the STAR results in the 0--10\% Ruthenium class \cite{STARquarkmatter} we show in Fig.~\ref{fig:RAAratios} the inclusive charged hadron $R_{\rm AA}$ versus $p_T$ for several centrality classes for Case 5\@. By construction Eqn.~\ref{eq:raa} reproduces the experimental results for the 0--10\% Ruthenium class, but the other points are results of our model. Even though a significant suppression is visible the ratio of \ru{} over \zr{} is close to unity. This can be explained by Fig.~\ref{fig:pathlengths}, where for central collisions the ratio of the temperature integral is also close to unity. For more peripheral collisions the difference is larger, but for those the $R_{\rm AA}$ is quite close to unity and the effect of energy loss is hence relatively modest by itself.

The \npart{} dependence of the nuclear modification factor for high $p_T$ hadrons is presented in Fig.~\ref{fig:RAAvsNpart}, in which the $R_{\rm AA}$ at the largest \npart{} (the last point in the figure) was again used for model calibration. 
\emph{Trajectum} agrees qualitatively and semi-quantitatively till about $N_\text{part}\sim20$, which corresponds to about 60\% in centrality. Indeed, data from higher centrality intervals are often difficult to describe both theoretically and also experimentally (see e.g.~\cite{Loizides:2017sqq, Huss:2020whe})\@.

Our main result is presented in Fig.~\ref{fig:yields}\@. Here, we show Ru/Zr ratios for both the $R_{\rm AA}$ and the total yield of hadrons for three representative nuclear shapes (cases 2, 3 and 5 respectively)\@. While the yield ratio is simply the product of the $R_{\rm AA}$ ratio with the \ncoll{} ratio as presented in Fig.~\ref{fig:ncoll} we see that the modifications of the yield are much stronger and also depend more strongly on the nuclear geometry. Importantly, the yields are also a more direct measurement that hence can have a smaller experimental uncertainty. We thus argue that yields of high $p_T$ hadrons can be a complementary observable sensitive to the nuclear structure of in particular isobars.

Even for our case 5 the ratios of the yields do not give a perfect description of the STAR data \cite{STARquarkmatter}\@. For centrality classes above about 40\% this is consistent with Fig.~\ref{fig:RAAvsNpart}, where it is seen that the model underestimates modifications for very peripheral collisions. At low $p_T$ the assumption of \ncoll{} scaling may break down due to hydrodynamic particle production from the QGP, and indeed it seems the model works better at higher $p_T$. We note that even though our model is fixed at the 0--10\% Ru centrality the Ru/Zr ratio at 0--10\% is still a non-trivial model result that works best for case 5.

\section{Discussion}

\begin{figure}[ht]
\includegraphics[width=0.35\textwidth]{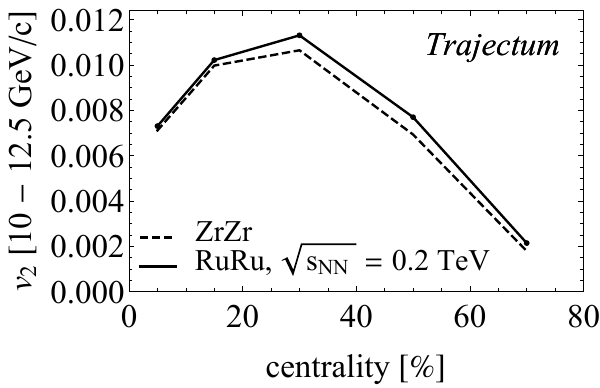}
\caption{\label{fig:v2highpt}Using the ratio of the in-plane versus out-plane path lengths together with the measured $R_{\rm AA}$ we can make an estimate of the elliptic flow $v_2$ as a function of $p_T$\@. We expect a small but significant anisotropic flow of up to 1.1\%.
}
\end{figure}

Given the in-plane and out-of-plane path lengths it is possible to compute the elliptic flow through the formula $v_2 \approx (1 - x)/(2 + 2 x)$, with $x = R_{\rm AA, in}/R_{\rm AA, out}$. The results are presented in Fig.~\ref{fig:v2highpt}, whereby interestingly Ruthenium has a significantly larger $v_2$. This can be traced back to the more anisotropic path lengths presented in Fig.~\ref{fig:pathlengths}. Interestingly, the difference in $v_2$ is largest in the 20--30\% centrality class, even though for the soft observables there is no difference in $v_2$ (see Appendix)\@.

One subtlety of the yield ratios as presented is that they are divided per centrality class. Since centrality is determined independently for Ru and Zr this gives a systematic uncertainty that does not cancel in the ratio. In this work we have taken equal \emph{Trajectum}-defined centrality classes and have not systematically estimated the uncertainty associated to that. On the other hand, computing the \ncoll{} ratio with the (differing) centrality classes for Ru and Zr as defined in \cite{STAR:2021mii} changes results by up to 0.5\% till 40\% centrality and by over 3\% beyond that (see Appendix)\@. For the STAR results in Fig.~\ref{fig:yields} it is not entirely clear how the STAR centralities are defined, but this could be a dominant uncertainty.

In summary, we have argued for hard probes in relativistic isobar collisions to investigate their nuclear structure and to provide a unique opportunity to study the behavior of nucleons at high energies and densities. \emph{Trajectum} calculations show that the number of binary collisions ($N_\text{coll}$) in RuRu and ZrZr collisions differs significantly in different centrality classes. The ratio of $N_\text{coll}$ depends sensitively on the shape of the nuclei. These make electroweak bosons, which provide direct access to the number of binary collisions in the isobar collisions, promising observables at the Large Hadron Collider. See also \cite{Zhao:2022mce} for a study using $J/\psi$ mesons, which due to regeneration can even be sensitive to \ncoll{}$^2$. Moreover, we demonstrated that due to the significant cancellation of jet quenching in QGP in the ratio, coloured hard probes, such as high momentum jets and charged particles that are more abundantly produced than electroweak bosons, could be used to achieve the same goal. The information obtained from these collisions has important implications for understanding neutron-rich nuclei, the quark-gluon plasma, and other phenomena in nuclear and astrophysics.

\section*{Acknowledgements}
We acknowledge discussions with participants of the INT Program INT-23-1a, "Intersection of nuclear structure and high‐energy nuclear collisions" and Institute for Nuclear Theory for its hospitality. GN is supported by the U.S. Department of Energy, Office of Science, Office of Nuclear Physics under grant Contract Number DE-SC0011090. YL and YC are supported by the U.S. Department of Energy, Office of Science, Office of Nuclear Physics under grant Contract Number DE-SC0011088.

\bibliography{main, manual}{}

\newpage

\section*{Appendix}

\begin{figure}[ht]
\includegraphics[width=0.5\textwidth]{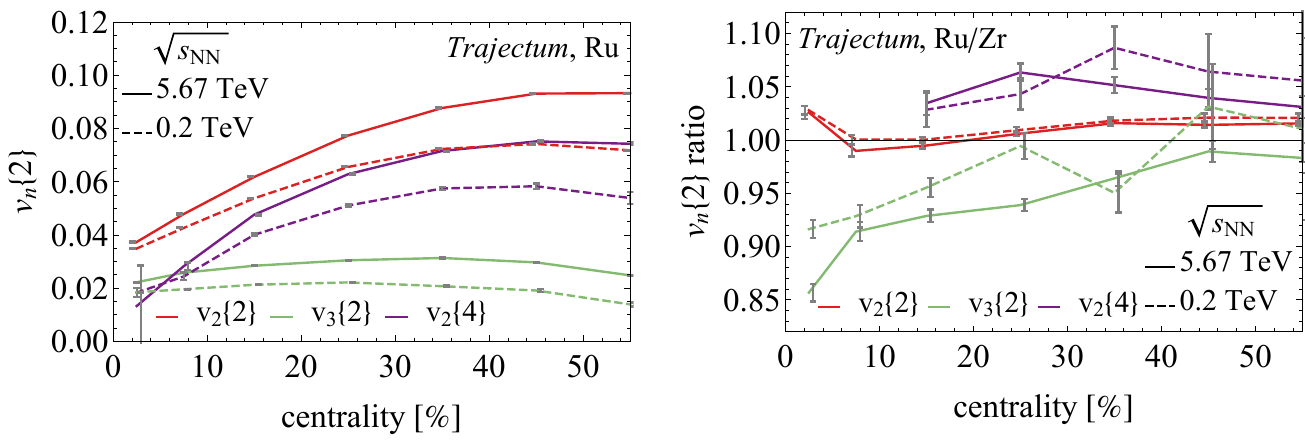}
\caption{\label{fig:v2at5}For Case 5 we present several anisotropic flow coefficients at 0.2\,TeV collision energy (see also \cite{Nijs:2021kvn}) and at 5.67\,TeV for Ru (left) together with the Ru/Zr ratio (right)\@. At higher energy we expect larger anisotropic flow signals with a stronger centrality dependence. It is interesting that especially the triangular flow ratio is significantly more modified at 5.67\,TeV\@.
}
\end{figure}

In Fig.~\ref{fig:v2at5} we show anisotropic flow coefficients from the hydrodynamic calculations for $0.2\,$TeV (reproduced from \cite{Nijs:2021kvn}) and $5.67\,$TeV\@. Interestingly, though the flow is larger at higher energy we see that the ratio is relatively robust even under this large change in collision energy. The exception is perhaps the triangular flow $v_3\{2\}$, which deviates more strongly for central collisions (see also \cite{Bhatta:2023cqf} for similar findings).

\begin{figure}[ht]
\includegraphics[width=0.5\textwidth]{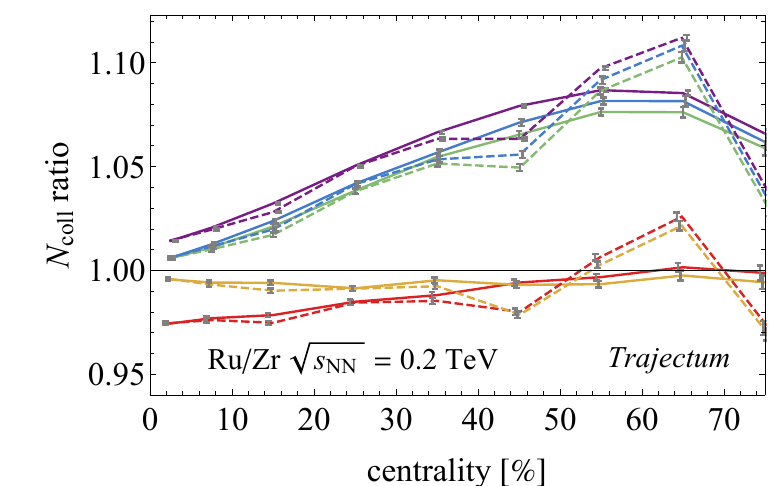}
\caption{\label{fig:ncollcentdefinition}We show the \ncoll{} ratios for different centrality definitions as done in \cite{STAR:2021mii} (dashed) as compared with the standard \emph{Trajectum} definition from Fig.~\ref{fig:ncoll} (solid)\@.
}
\end{figure}

In Fig.~\ref{fig:ncollcentdefinition} we compare our standard centrality definition for \ncoll{} with the STAR definition from \cite{STAR:2021mii}\@. In the STAR definition Ru and Zr centrality classes are slightly different to ease the  experimental analysis. We see that especially for peripheral collisions this is an important effect (similar to the effect on the multiplicity, see \cite{Nijs:2021kvn}), so that clarity on the precise centrality selection will be essential for a precision analysis.

\end{document}